\documentstyle[preprint,aps,epsfig]{revtex}
\newcommand{\gsim}{\mathrel{\rlap{\lower4pt\hbox{\hskip0pt$\sim$}}
\raise1pt\hbox{$>$}}}
\newcommand{\lsim}{\mathrel{\rlap{\lower4pt\hbox{\hskip0pt$\sim$}}
\raise1pt\hbox{$<$}}}
\tighten
\begin{document}
%
\preprint{Preprint Number: \parbox[t]{45mm}{ANL-PHY-9350-TH-99\\
}}
 \title{Describing $a_1$ and $b_1$ decays}
\author{J. C. R. Bloch,\footnotemark[1] 
        Yu. L. Kalinovsky,\footnotemark[2] 
        C. D. Roberts\footnotemark[1] and
        S. M. Schmidt\footnotemark[1]\vspace*{0.2\baselineskip}}

 \address{
 \footnotemark[1]Physics Division, Bldg. 203, Argonne National Laboratory,
 Argonne IL 60439-4843\vspace*{0.2\baselineskip}\\
\footnotemark[2]Laboratory of Computing Techniques and Automation,
 \\Joint Institute for Nuclear Research, 141980 Dubna,
 Russia\vspace*{0.2\baselineskip}}  
\date{Pacs Numbers: 13.25.-k, 14.40.Cs, 24.85.+p, 12.38.Lg}
\maketitle
\begin{abstract}
Two-body pion-radiating and weak decays of light axial-vector mesons and the
$\rho$ are studied as a phenomenological application of the QCD
Dyson-Schwinger equations.  Models based on the rainbow-ladder truncation are
capable of providing a good description and, in particular, yield the correct
sign and magnitude of the $a_1\to\rho\pi$ and $b_1\to\omega\pi$ $D/S$ ratios,
with no additional mechanism necessary.
\end{abstract}
The light meson spectrum contains\cite{pdg} four little-studied axial-vector
mesons composed of $u$- and $d$-quarks.  They appear as isospin $I=0,1$
partners (in the manner of the $\omega$ and $\rho$): $h_1(1170)$,
$b_1(1235)$; and $f_1(1285)$, $a_1(1260)$, and differ in their charge-parity:
$J^{PC}=1^{+-}$ for $h_1$, $b_1$; and $J^{PC}= 1^{++}$ for $f_1$, $a_1$.  In
the $q\bar q$ constituent quark model the $b_1$ is represented as a
constituent-quark and -antiquark with total spin $S=0$ and angular momentum
$L=1$, while in the $a_1$ the quark and antiquark have $S=1$ and $L=1$.  It
is therefore apparent that in this model the $b_1$ is an orbital excitation
of the $\pi$, and the $a_1$ an orbital excitation and axial-vector partner of
the $\rho$.  In QCD the $J^{PC}$ characteristics of a quark-antiquark bound
state are manifest in the structure of its Bethe-Salpeter
amplitude\cite{ls69}.  This amplitude is a valuable intuitive guide and, in
cases where a $q\bar q$ constituent quark model analogue exists, it
incorporates and extends the information present in that analogue's quantum
mechanical wave function.  We describe mesons via their Bethe-Salpeter
amplitudes.

Three of the axial-vector mesons decay predominantly into two-body final
states containing a vector meson and a pion:\footnote{
$f_1 \to \rho\pi$ is forbidden by $G$-parity conservation and $f_1\to \omega
\pi$ does not conserve isospin.  Therefore $f_1$ decays are dominated by
four-pion final states, which makes them harder to employ as an intuition
building tool.}
$h_1\to \rho\pi$; $b_1\to \omega \pi$; $a_1\to \rho\pi$, and with a $J=1$
meson in both the initial and final state they proceed via two partial waves
($S$, $D$).  Such decays therefore probe aspects of hadron structure
inaccessible in simpler processes involving only spinless mesons in the final
state, such as $\rho\to\pi\pi$; e.g., in constituent-quark-like models the
$D/S$ amplitude ratio is very sensitive\cite{eric} to the nature of the
phenomenological long-range confining interaction.  The additional insight
and model constraints that such processes can provide is particularly
important now as a systematic search and classification of ``exotic''
states\footnote{
A meson is labelled\cite{pdg} ``exotic'' if it is characterised by a value of
$J^{PC}$ that is unobtainable in the $q\bar q$ constituent quark model; e.g.,
the $1.6\,$GeV $J^{PC}=1^{-+}$ state for which experimental
evidence\cite{exotic} has recently appeared.  Such unusual charge parity
states are a necessary feature of a field theoretical description of
quark-antiquark bound states\cite{ls69} with Bethe-Salpeter equation studies
typically yielding\cite{bsesep} masses approximately twice as large as that
of the natural charge parity partner and, in particular, a $J^{PC}=1^{-+}$
meson with a mass\cite{bpprivate} of $\sim 1.4$ -- $1.5\,$GeV. }
in the light meson sector becomes feasible experimentally.  Therefore, as an
illustration of the application\cite{peterrev} of the QCD Dyson-Schwinger
equations (DSEs)\cite{cdragw}, and an exploration and further elucidation of
the domain of applicability of commonly used truncations, we report a
simultaneous study of these axial-vector meson decays, the $\rho\to\pi\pi$
decay, and the leptonic decay constants.  The breadth of the study ensures a
global view.

In the isospin symmetric limit the homogeneous Bethe-Salpeter equation (BSE)
for a quark-antiquark bound state is
\begin{equation}
\label{bsegen}
\Gamma^{tu}(k;P)= \int_q^\Lambda\,\chi^{sr}(q;P)\,K^{rs}_{tu}(q,k;P),
\end{equation}
where $k$ is the relative and $P$ the total momentum of the constituents,
$\chi(q;P):= S(q_+)\Gamma(q;P)S(q_-)$, $r,\ldots,u$ represent colour, Dirac
and isospin indices, $q_{\pm}= q\pm P/2$, and
$\int_q^\Lambda:= \int^\Lambda d^4q/(2\pi)^4 $
represents mnemonically a translationally invariant regularisation of the
integral, with $\Lambda$ the regularisation mass-scale.  In
Eq.~(\ref{bsegen}), $S$ is the renormalised dressed-quark propagator and $K$
is the renormalised, fully-amputated dressed-quark-antiquark scattering
kernel.  The equation has a solution $\Gamma$, the Bethe-Salpeter amplitude,
only for particular, isolated values of $P^2$, which determine the mass of
the associated bound state.  The amplitude is a necessary element in the
calculation of any of the meson's interactions.

The renormalised dressed-quark propagator in Eq.~(\ref{bsegen}) is determined
by the quark DSE\footnote{We use a Euclidean formulation with
$\{\gamma_\mu,\gamma_\nu\}=2\delta_{\mu\nu}$, $\gamma_\mu^\dagger =
\gamma_\mu$, $p\cdot q=\sum_{i=1}^4 p_i q_i$, and
tr$_D[\gamma_5\gamma_\mu\gamma_\nu\gamma_\rho\gamma_\sigma]=
-4\,\epsilon_{\mu\nu\rho\sigma}$, $\epsilon_{1234}= 1$.  A vector, $k_\mu$,
is timelike if $k^2<0$.}
\begin{eqnarray}
\label{genS}
S(p)^{-1} & := & 
i \gamma\cdot p \,A(p^2) + B(p^2)  \\
%
%
\label{gendse} & = & Z_2 (i\gamma\cdot p + m^{\rm bm})
+ Z_1\, \int^\Lambda_q \,
\case{4}{3} \,g^2 D_{\mu\nu}(p-q) \gamma_\mu S(q)\Gamma_\nu(q,p),
\end{eqnarray}
where $D_{\mu\nu}(k)$ is the renormalised dressed-gluon propagator,
$\Gamma_\nu(q,p)$ is the renormalised dressed-quark-gluon vertex, and $m^{\rm
bm}$ is the $\Lambda$-dependent $u$-, $d$-current-quark bare mass.  The
renormalisation constants: $Z_1(\zeta^2,\Lambda^2)$ and
$Z_2(\zeta^2,\Lambda^2)$, depend on the renormalisation point, $\zeta$, and
the regularisation mass-scale.

To proceed one must identify a reliable truncation of the integral kernels in
Eqs.~(\ref{bsegen}) and (\ref{gendse}).  Qualitatively reliable results are
unobtainable unless the truncations ensure the preservation of the
Ward-Takahashi identities that relate the dressed-quark propagator to the
solution of the relevant inhomogeneous Bethe-Salpeter equations.  Satisfying
this constraint, results as important as the correlation between dynamical
chiral symmetry breaking and the low mass of the pion (Goldstone's theorem)
are automatic\cite{pct98}.  For most light mesons\footnote{
The exceptions are the $\eta$-$\eta^\prime$ complex and light mesons that
have vacuum quantum numbers; i.e., $J^{PC}= 0^{++}$, for which higher orders
must be retained in the expansion of $\Gamma_\nu$ and $K$ introduced and
explored in Refs.\protect\cite{brs96}.}
the class of renormalisation-group-improved rainbow-ladder truncations is
satisfactory.  It is reliable in Landau gauge, a fixed point of the QCD
renormalisation group, and defined by
\begin{eqnarray}
\label{Ga}
Z_1 \,g^2 D_{\mu\nu}(k-q)\,\Gamma_\nu(q,p) & = &
{\cal G}(k-q) \,D^{\rm free}_{\mu\nu}(k-q) \,\gamma_\nu,\\
\label{Gb}
K^{rs}_{tu}(q,k;P) & = & - {\cal G}(k-q) D^{\rm free}_{\mu\nu}(k-q)\,
\left(\case{1}{2}\lambda^a\gamma_\mu\right)_{tr}
\left(\case{1}{2}\lambda^a\gamma_\nu\right)_{su},
\end{eqnarray}
with $D^{\rm free}_{\mu\nu}(k)= (\delta_{\mu\nu}-k_\mu k_\nu/k^2)/k^2$ and
${\cal G}(k^2)/(4\pi) = \alpha(k^2)$ for $k^2 \gsim 1\,$GeV$^2$.  This
truncation ensures that the solutions of the DSEs exhibit that momentum
evolution characteristic of the QCD renormalisation group at one-loop order.
For $k^2\lsim 1\,$GeV$^2$, the form of ${\cal G}(k^2)$ is only loosely
constrained and in contemporary phenomenological studies it is modelled,
often in accordance with the constraints\cite{hawes} supplied by studies of
the gluon DSE\cite{cdragw,mrp}.

An early, extensive study of the light meson spectrum using
Eq.~(\ref{bsegen}) in rainbow-ladder truncation is reported in
Ref.\cite{pankaj}.  It neglected axial-vector mesons.  Improved studies,
accounting for the complete Dirac structure of pseudoscalar\cite{mr97} and
vector\cite{mt99} meson Bethe-Salpeter amplitudes and explaining many of the
important contributions made by the nonleading components, such as the
pseudovector component of the pion, have been completed, but the axial-vector
mesons still await such a careful treatment.  In its absence we use a crude,
confining, Goldstone theorem preserving rank-2 separable Ansatz\cite{bsesep}
for ${\cal G}(k^2)$ to provide guidance to the structure of the light mesons.

The foundation of the separable Ansatz is a model\cite{cdrpion} dressed-quark
propagator:
\begin{eqnarray}
\label{defS}
S(p) & = & -i\gamma\cdot p\,\sigma_V(p^2) + \sigma_S(p^2) ,\\
\label{SSM}
\bar\sigma_S(x)  & =  & 2 \bar m_f {\cal F}(2 (x + \bar m_f^2))
        + {\cal F}(b_1 x) {\cal F}(b_3 x) 
        \left[ b_0 + b_2 {\cal F}(\epsilon x)\right]\,,\\
\label{SVM}
\bar\sigma_V(x) & = & \frac{1}{x + \bar m_f^2}
        \left[ 1 - {\cal F}(2 (x+\bar m_f^2))\right] = 
        \frac{ 2 (x+\bar m_f^2) - 1 + {\rm e}^{-2 (x+\bar m_f^2)}}
        {2 (x+\bar m_f^2)^2}
\end{eqnarray}
with: $\bar\sigma_S(x) = \lambda\,\sigma_S(p^2)$, $\bar\sigma_V(x) =
\lambda^2\,\sigma_V(p^2)$; ${\cal F}(y)= (1-{\rm e}^{-y})/y$;
$x=p^2/\lambda^2$; $\bar m_f$ = $m_f/\lambda$; and $\lambda$ a mass scale.
This confining model efficiently characterises many essential and robust
elements of the solution of Eq.~(\ref{gendse}), and has been used
efficaciously\cite{peterrev} in a wide range of phenomenological
applications: most recently in a survey of heavy meson
observables\cite{mishaPRD} and an elucidation\cite{pwc99} of the effect of
meson loops on $\rho$-meson properties.

In basing the separable Ansatz on this algebraic model, renormalisability is
lost and a regularisation must be introduced to properly define the
Dyson-Schwinger and Bethe-Salpeter equations.  It is sufficient\cite{bsesep}
to modify $\sigma_S$ and $\sigma_V$ so that
$ {\cal F}(b_1 x) \to {\cal F}((\epsilon_s x)^2) {\cal F}(b_1 x) $
in Eq.~(\ref{SSM}) and 
$ [2 (x+\bar m^2) - 1] \to [2 (x+\bar m^2) - \exp(-\epsilon_V^2 (x+\bar m^2))]$
in Eq.~(\ref{SVM}).  The parameters take the values
\begin{equation}
\label{params}
\begin{array}{ccccccc}
\bar m & b_0 & b_1 & b_2 & b_3 & \epsilon_S & \epsilon_V \\
0.00811 & 0.131 & 2.90 & 0.603 & 0.185 & 0.482 & 0.1 
\end{array}
\end{equation}
with $\lambda = 0.566\,$GeV, and because it is a separable Ansatz the meson
Bethe-Salpeter amplitudes are expressed solely in terms of two
functions:\footnote{ 
Detailed studies\protect\cite{pankaj,mr97,mt99} show that this is flawed;
i.e., in general, the momentum dependence of the modulating functions in the
Bethe-Salpeter amplitudes is {\bf not} related to the scalar functions in the
quark propagator in such a trivial fashion.  However, we only require that
the separable Ansatz provide initial guidance and herein also report results
obtained with a rudimentary amelioration of this defect.}
\begin{equation}
\label{FG}
\bar F(x):= \case{1}{a}\left[\bar A(x)-1\right],\;
\bar G(x):= \case{1}{b}\left[\bar B(x)-\bar m\right],
\end{equation}
with calculated values of $a= 0.129 $, $b=0.0877 $.  

We can now solve the Bethe-Salpeter equation to obtain the masses and
amplitudes of the participating mesons: the light axial-vectors, and the
$\pi$, $\rho$ and $\omega$.  Using the separable Ansatz the general form of
their Bethe-Salpeter amplitudes is
\begin{eqnarray}
\label{gammapi}
\Gamma^\pi(\bar k;\hat P) & = & \vec{\tau}\,
\gamma_5 \left(i e^\pi_1 - e^\pi_2 \gamma\cdot \hat P \right)\bar G(x), \\
\Gamma_\mu^\rho(\bar k;\hat P) & = & \vec{\tau}\,\left(
\bar k^T_\mu e_1^\rho\, \bar F(x)
+ i e_2^\rho \gamma_\mu^T \,\bar G(x)
- i e_3^\rho \gamma_5 
  \epsilon_{\lambda\mu\nu\sigma}\gamma_\lambda \bar k_\nu \hat P_\sigma \,
                \bar F(x)        \right),\\
\Gamma_\mu^{b_1}(\bar k;\hat P) & = & \vec{\tau}\,\bar k^T_\mu \, 
\gamma_5 \left(i e_1^{b_1} - e_2^{b_1} \gamma\cdot \hat P\right)\bar F(x),\\
\Gamma_\mu^{a_1}(\bar k;\hat P) & = & \vec{\tau} \, \left(
        i\gamma_5\gamma_\mu^T e_1^{a_1} \bar G(x)
 + ie_2^{a_1}\,
        \epsilon_{\lambda\mu\nu\sigma}\gamma_\lambda \bar k_\nu \hat P_\sigma 
        \bar F(x) \right),
\end{eqnarray}
where $\bar k = k/\lambda$, $\hat P_\mu = P_\mu/|P^2|^{1/2}$, $\bar k_\mu^T =
(\bar k_\mu + \hat P_\mu \bar k\cdot \hat P)$, and $\gamma_\mu^T =
(\gamma_\mu + \hat P_\mu \gamma\cdot\hat P)$.  Here the eigenvectors
$\vec{e}^{\,H}$, obtained by solving the Bethe-Salpeter equation, are all
that remain in the separable Ansatz of the detailed structure exhibited by a
realistic Bethe-Salpeter amplitude.  Qualitative information nevertheless
remains; e.g., comparing the $\pi$ and $b_1$ amplitudes it is apparent that
the $b_1$ has characteristics consistent with an orbital excitation.
Further, even using this Ansatz one observes that the momentum dependence of
the scalar functions modulating the amplitudes is very different: $\bar G$
for $\pi$, $\bar F$ for $b_1$, a sharp contrast with
constituent-quark-model-like treatments.  The Bethe-Salpeter amplitudes of
the $\omega$, $h_1$, $f_1$ are obtained from those of the $\rho$, $b_1$,
$a_1$ via the replacement $\vec{\tau} \to \tau^0:= \mbox{\rm diag}(1,1)$.
These partners are degenerate in rainbow-ladder truncation
(cf. experimentally, none of the mass differences is greater than 5\% of the
average mass).

The separable Bethe-Salpeter equation is easily solved and yields the
following masses and eigenvectors (quoted here as normalised trivially via
$\Sigma e_i^2 = 1$)
\begin{equation}
\label{sepout}
\begin{array}{l|ll|rrr}
        & m_{\rm exp} ({\rm GeV}) & m ({\rm GeV}) &
                       e_1   & e_2   &       e_3 \\\hline
\pi        & 0.139& 0.139     & 0.9973     & -0.0732 &       \\
\omega,\rho& 0.77 & 0.736     & -0.2270     &  0.9610 & -0.1580 \\
h_1,b_1 & 1.23 \pm 0.003 & 1.24 & 0.9708     & 0.2400 &       \\
f_1,a_1 & 1.23 \pm 0.040& 1.34 & 0.1991      & 0.9800 &
\end{array}
\end{equation}
thus reproducing some of the results in Ref.\cite{bsesep} and providing the
first calculation of the $b_1$.  (The observed\cite{pdg} isovector masses are
quoted for comparison.)  Here, as with variational wave functions obtained in
the estimation of energy levels, an accurate value of the mass does not
necessarily mean that the model Bethe-Salpeter amplitude will be reliable for
the calculation of transitions.  Nevertheless, the suggestions about the form
cannot be ignored, and it is interesting and important to observe that the
$\gamma_5\gamma_\mu$ term in the $a_1$ amplitude is {\bf not} the dominant
contribution, which is provided instead by the second term that exhibits
characteristics of orbital motion.

The correct, canonical normalisation of the Bethe-Salpeter amplitude ensures
that the pole associated with the meson in the quark-antiquark scattering
amplitude: $M:= K + K (SS) K + \ldots$, has unit residue.  In calculations of
observables one must therefore rescale the naively normalised eigenvectors in
Eq.~(\ref{sepout}) by a constant factor: $e_i^H\to e_i^H/N_H$, which for the
pion is obtained in rainbow-ladder truncation from 
\begin{eqnarray}
\label{Npi}
\lefteqn{2 N_\pi^2 \,\delta^{ij}\,P_\mu = }\\
\nonumber && {\rm tr} \int^\Lambda_q 
\left.\left[ \Gamma^{\pi i}(\bar q;-\hat P) 
        \frac{\partial S(q_+)}{\!\!\!\!\!\!\partial P_\mu} 
        \Gamma^{\pi j}(\bar q;\hat P) S(q_-)  + 
\Gamma^{\pi i}(\bar q;-\hat P) S(q_+) \Gamma^{\pi j}(\bar q;\hat P) 
        \frac{\partial S(q_-)}{\!\!\!\!\!\!\partial P_\mu}\right]
        \right|_{P^2=-m_\pi^2}\! ,
\end{eqnarray}
where $i,j=1,2,3$ label the Pauli matrices in Eq.~(\ref{gammapi}) and the
trace is over colour, Dirac and isospin indices, and for the $J=1$ mesons
from ($\alpha,\beta = 0,1,2,3$)
\begin{eqnarray}
\label{NH}
\lefteqn{2 N_H^2 \,\delta^{\alpha\beta} \, P_\mu = }\\
&& 
\nonumber
\left.\case{1}{3}{\rm tr} \int^\Lambda_q 
\left[ \Gamma_\nu^{H \alpha}(\bar q;-\hat P) 
        \frac{\partial S(q_+)}{\!\!\!\!\!\!\partial P_\mu} 
        \Gamma_\nu^{H \beta}(\bar q;\hat P) S(q_-)  + 
\Gamma_\nu^{H \alpha}(\bar q;-\hat P) S(q_+) \Gamma_\nu^{H \beta}(\bar q;\hat P) 
        \frac{\partial S(q_-)}{\!\!\!\!\!\!\partial P_\mu}\right]
        \right|_{P^2=-m_H^2}\! .
\end{eqnarray}

Our primary focus is the pionic decays of the axial-vector mesons.  The decay
$a_1(P_\mu) \to \rho(q_\nu) \pi(k)$, $P^2=-m_{a_1}^2$, $q^2=-m_\rho^2$,
$k^2=-m_\pi^2$, is described by an amplitude
\begin{eqnarray}
\label{Taropi}
T^{a_1\rho\pi}_{\mu\nu} & = & {\cal A}(m_{a_1}^2,m_\rho^2)\,G_{\mu\nu} +
{\cal B}(m_{a_1}^2,m_\rho^2)\,L_{\mu\nu}, 
\end{eqnarray}
that we have expressed in terms of two projection operators ($Y = (P\cdot
q)^2 - m_{a_1}^2 m_\rho^2$):
\begin{eqnarray}
G_{\mu\nu} & = & \delta_{\mu\nu} 
- \frac{1}{Y} \left[ m_{a_1}^2 q_\mu q_\nu + m_\rho^2 P_\mu P_\nu
                + P\cdot q \left( P_\mu q_\nu + q_\mu P_\nu\right)\right],\\
L_{\mu\nu} & = & \frac{P\cdot q}{Y}
\left(P_\mu + q_\mu \frac{m_{a_1}^2}{P\cdot q}\right)
\left(q_\nu + P_\nu \frac{m_{\rho}^2}{P\cdot q}\right),
\end{eqnarray}
with the properties: $P_\mu G_{\mu\nu}= 0 = G_{\mu\nu} q_\nu$, $P_\mu
L_{\mu\nu}= 0 = L_{\mu\nu} q_\nu$, a manifestation of the on-shell
transversality of $J=1$ mesons; $G_{\mu\nu}G_{\mu\nu}=2$; $L_{\mu\nu}
L_{\mu\nu} = m_{a_1}^2 m_\rho^2/(P\cdot q)^2$; and $G_{\mu\nu}L_{\mu\nu}=0$.
The scalar functions: ${\cal A}(m_{a_1}^2,m_\rho^2)$ and ${\cal
B}(m_{a_1}^2,m_\rho^2)$, contain all the dynamical information about this
process; e.g., from Eq.~(\ref{Taropi}) one obtains
\begin{equation}
\Gamma_{a_1\rho\pi}= \frac{1}{12\pi}\frac{|\vec{k}|}{m_{a_1}^2}
        \left(2 |{\cal A}|^2 
        + \frac{ m_{a_1}^2 m_\rho^2 }{(P\cdot q)^2} |{\cal B}|^2\right),
\end{equation}
with $|\vec{k}|^2= \lambda(m_{a_1}^2,m_\rho^2,m_\pi^2)/(2m_{a_1})^2$, 
$\lambda(m_{a_1}^2,m_\rho^2,m_\pi^2)=
(m_{a_1}^2-(m_\rho+m_\pi)^2) (m_{a_1}^2-(m_\rho-m_\pi)^2)$,
and following Ref.~\cite{isgur} the ratio
\begin{equation}
\label{dons}
\left.D/S\right|_{a_1 \rho \pi} =
\frac{f^D_{a_1 \rho \pi}}
{f^S_{a_1 \rho \pi}}
\end{equation}
where
\begin{eqnarray}
f^S_{a_1 \rho \pi}(m_{a_1}^2,m_\rho^2) & = & 
\frac{\sqrt{4\pi}}{3m_\rho}\left[
(E_\rho + 2m_\rho) f_{a_1 \rho \pi}(m_{a_1}^2,m_\rho^2)
+ |\vec{k}|^2 m_{a_1} g_{a_1 \rho \pi}(m_{a_1}^2,m_\rho^2)\right],\\
\label{fD}
f^D_{a_1 \rho \pi}(m_{a_1}^2,m_\rho^2) & = & 
-\frac{\sqrt{8 \pi}}{3m_\rho}\left[
(E_\rho - m_\rho) f_{a_1 \rho \pi}(m_{a_1}^2,m_\rho^2)
+ |\vec{k}|^2 m_{a_1} g_{a_1 \rho \pi}(m_{a_1}^2,m_\rho^2)\right],
\end{eqnarray}
with $E_\rho^2 = |\vec{k}|^2 + m_\rho^2 $,
$f_{a_1 \rho \pi}(m_{a_1}^2,m_\rho^2) = {\cal A}(m_{a_1}^2,m_\rho^2)$;
and
\begin{equation}
g_{a_1 \rho \pi}(m_{a_1}^2,m_\rho^2)=
\frac{P\cdot q}{Y}\left(-{\cal A}(m_{a_1}^2,m_\rho^2) 
     + {\cal B}(m_{a_1}^2,m_\rho^2)
        \frac{m_{a_1}^2 m_\rho^2}{(P\cdot q)^2}\right).
\end{equation}

We calculate $T_{\mu\nu}^{a_1\rho\pi}$ in impulse
approximation:$\,$\footnote{The analogous approximation to the matrix element
for the $f_1 \to\,$vector-meson-plus-pion decay is easily shown to yield
${\cal A}\equiv 0 \equiv {\cal B}$.}
\begin{equation}
\label{a1imp}
iT_{\mu\nu}^{a_1\rho\pi}(P,q)=  \,
2 \,{\rm tr}\int_\ell^\Lambda\,
S(\ell_{--}) \Gamma_\mu^{a_1}(\bar \ell;-\hat P)
S(\ell_{++}) \Gamma^\pi(\bar\ell_{+0};\hat k)
S(\ell_{+-}) \Gamma^\rho_\nu(\bar\ell_{0-};\hat q)
\end{equation}
where $\ell_{\alpha\beta}= \ell + \alpha q/2 + \beta k/2$.  It is the
simplest approximation consistent with the rainbow-ladder truncation and
corrections can be incorporated systematically\cite{brs96}.
Calculations\cite{impulse} indicate that they contribute $\lsim 15$\% for
on-shell momenta.  With our estimate of the meson Bethe-Salpeter amplitudes
and form for the dressed-quark propagator, Eq.~(\ref{a1imp}) is
straightforward to evaluate, remembering that here and below the
Bethe-Salpeter amplitudes are rescaled by $N_H$, as discussed in connection
with Eqs.~(\ref{Npi}) and (\ref{NH}).

The $b_1(P)\to \omega(q) \pi(k)$ matrix element also has the representation
in Eq.~(\ref{Taropi}) and in impulse approximation is
\begin{equation}
\label{Tb1opi}
T_{\mu\nu}^{b_1\omega\pi}(P,q)=  \,
2 \,{\rm tr}\int_\ell^\Lambda\,
S(\ell_{--}) i\Gamma_\mu^{b_1}(\bar \ell;-\hat P)
S(\ell_{++}) i\Gamma^\pi(\bar\ell_{+0};\hat k)
S(\ell_{+-}) \Gamma^\omega_\nu(\bar\ell_{0-};\hat q),
\end{equation}
with the width given by
\begin{equation}
\Gamma_{b_1\omega\pi} =  \frac{1}{24\pi}\frac{|\vec{k}|}{m_{b_1}^2}
        \left(2 |{\cal A}(m_{b_1}^2,m_\omega^2)|^2 
        + \frac{ m_{b_1}^2 m_\omega^2 }{(P\cdot q)^2} 
        |{\cal B}(m_{b_1}^2,m_\omega^2)|^2\right),
\end{equation}
$|\vec{k}|^2= \lambda(m_{b_1}^2,m_\omega^2,m_\pi^2)/(2m_{b_1})^2$, and
$D/S|_{b_1\omega\pi}$ obtained by analogy with Eqs.~(\ref{dons})-(\ref{fD}).

$h_1(P)\to \rho(q) \pi(k)$ is directly accessible too, with the impulse
approximation to its matrix element obtained via the obvious modifications of
Eq.~(\ref{Tb1opi}): $b_1\to h_1$, $\omega\to\rho$, the width given by
\begin{equation}
\Gamma_{h_1\rho\pi} =  \frac{1}{8\pi}\frac{|\vec{k}|}{m_{h_1}^2}
        \left(2 |{\cal A}(m_{h_1}^2,m_\rho^2)|^2 
        + \frac{ m_{h_1}^2 m_\rho^2 }{(P\cdot q)^2} 
        |{\cal B}(m_{h_1}^2,m_\rho^2)|^2\right),
\end{equation}
$|\vec{k}|^2= \lambda(m_{h_1}^2,m_\rho^2,m_\pi^2)/(2m_{h_1})^2$, and
$D/S|_{h_1\rho\pi}$ obtained as the obvious analogue of Eq.~(\ref{dons}).

Our results for these strong decays can be placed in perspective by comparing
them with those for the $\rho(P)\to\pi(q)\pi(k)$ decay, whose matrix element
can be written
\begin{equation}
M_\mu(k,q)= (k-q)_\mu f^+(t) + P_\mu f^-(t),
\end{equation}
$t= -(k-q)^2$, in which case 
\begin{equation}
\Gamma_{\rho\pi\pi} = \frac{g_{\rho\pi\pi}^2}{6\pi} 
        \frac{|\vec{k}|^3}{m_\rho^2},\;
g_{\rho\pi\pi}= \case{1}{2} f^+(t=m_\rho^2),
\end{equation}
$f^-(m_\rho^2)\equiv 0$,
$|\vec{k}|^2=\lambda(m_{\rho}^2,m_\pi^2,m_\pi^2)/(2m_{\rho})^2$.  The impulse
approximation to $M_\mu(k,q)$ is
\begin{equation}
\label{Mrpp}
iM_{\mu}(k,q)=  \,
{\rm tr}\int_\ell^\Lambda\,
S(\ell_{--}) \Gamma_\mu^{\rho}(\bar \ell;-\hat P)
S(\ell_{++}) \Gamma^\pi(\bar\ell_{+0};\hat k)
S(\ell_{+-}) \Gamma^\pi(\bar\ell_{0-};\hat q).
\end{equation}

This perspective is sharpened by also calculating the meson decay constants,
which for $J=1$ are given in QCD by\cite{mishaPRD}
\begin{equation}
\label{fH}
i\sqrt 2 \,\delta^{\alpha\beta}\,f_H \,m_H  = 
\case{1}{3}{\rm tr}Z_2\int_k^\Lambda\,
(\gamma_\mu - \gamma_\mu\gamma_5)\tau^\alpha 
        S(k_+)\Gamma_\mu^{H\beta}(\bar k;\hat P)S(k_-).
\end{equation}
These decay constants are the analogue of those associated with pseudoscalar
mesons, the expression for which is given in Ref.\cite{mr97}, and completely
describe\cite{mt99,mishaPRD} the strong interaction contribution to the weak
decays of the charged mesons and electromagnetic decays of the
neutral.\footnote{
The factor $Z_2(\zeta^2,\Lambda^2)$ in Eq.~(\ref{fH}) ensures that the
right-hand-side is finite as $\Lambda\to\infty$ and is renormalisation-point
and gauge-parameter independent.  The manner in which this is realised in QCD
is nevertheless interesting, requiring a conspiracy between various
components of the meson Bethe-Salpeter amplitudes, and is elucidated in
Ref.\protect\cite{mt99}.}
We note that because of the charge-parity of the $h_1$, $b_1$ mesons, which
is manifest in the Dirac component of their Bethe-Salpeter amplitudes via:
\begin{equation}
\bar\Gamma^H_\mu(\bar k;\hat P):= 
\left(C^{-1} \Gamma^H_\mu(-\bar k;\hat P) C\right)^{T}
= - \Gamma^H_\mu(\bar k;\hat P),
\end{equation}
$C=\gamma_2\gamma_4$ is the charge-conjugation matrix, it follows as a
model-independent result from Eq.~(\ref{fH}) that
\begin{equation}
f_{h_1} \equiv 0 \equiv f_{b_1}.
\end{equation}
This simply reflects the fact that a $V-A$ operator cannot connect a $1^{+-}$
state to the $0^{++}$ vacuum.

For the first of our calculations all the necessary elements are now defined:
the model dressed-quark propagator, Eqs.~(\ref{defS})-(\ref{params}); and the
Bethe-Salpeter amplitudes, Eqs.~(\ref{FG})-(\ref{NH}).  With these
amplitudes, which alone are regularised as described before
Eq.~(\ref{params}), we have a simple model in which the meson decay constants
are finite and $Z_2\to 1$.  A direct calculation yields the results in column
I of Table~\ref{tablea}.  Superficially: the decay constants are broadly
acceptable, the $D/S$ ratios have the correct sign, and the widths are too
large, resulting from an overestimate of the couplings ${\cal A}$, ${\cal
B}$, $g_{\rho\pi\pi}$ by a factor of $\sim 1.5$--$3.0$.  Delving further we
find that the results are very sensitive to the eigenvectors in
Eq.~(\ref{sepout}) and the form of the scalar functions that modulate the
different components of the Bethe-Salpeter amplitude.  These are
model-dependent features that are determined by the infrared behaviour of the
interaction; i.e., the form of ${\cal G}(k^2)$ for $k^2\lsim 1\,$GeV$^2$.  In
a sense this is in qualitative agreement with the observation\cite{eric}
quoted in the introduction; i.e., that the $D/S$ ratios are sensitive to the
long-range form of the quark-antiquark interaction.

It is important to observe that the values of the $D/S$ ratios and the ratio
of these ratios demonstrate that the class of rainbow-ladder DSE models
should {\bf not} be confused with the class of one-gluon exchange models used
extensively in constituent-quark-like models of hadrons, which necessarily
yield\cite{eric} the {\bf wrong} sign for the $D/S$ ratios.  It is incorrect
to anticipate defects in the former based on those of the latter.  We are
able to obtain the wrong sign for the $D/S$ ratios; e.g., forcing
$e_1^{a_1}=0$ with no other modification yields: $D/S|_{a_1\rho\pi}= 0.27$,
$\Gamma_{a_1\rho\pi} = 0.141\,$GeV and $f_{a_1}=0.154\,$GeV.  However, since
the Bethe-Salpeter amplitudes are dynamically determined this simply confirms
that the ratios do indeed provide useful constraints.  This sensitivity to
details of the Bethe-Salpeter amplitude is another example of that observed
in Ref.\cite{bsesep} and emphasised by the results described in
Refs.\cite{pmpion}.

An obvious question arises: is there a model for the scalar function ${\cal
G}(k^2)$ that can provide a good description of the observables; i.e., are
these phenomena describable using the rainbow-ladder truncation of the DSEs
with no additional mechanism?  The success of studies such as
Refs.~\cite{mr97,mt99} suggest that they are.  Here we follow a standard
approach and address a simpler question.  Since ${\cal G}(k^2)$ determines
the dressed-quark propagator and the Bethe-Salpeter amplitudes, then
modelling these elements instead provides a {\it de facto} model of ${\cal
G}(k^2)$: after all, that is the rationale behind separable
Ans\"atze\cite{bsesep,gunner}.  To relax the constraints imposed by the
rank-2 separable Ansatz we allow $b_2^{+}:=b_2^{a_1}=b_2^{b_1}$, $b_2^\pi$,
$b_0^\rho$, $b_1^\rho$ and $e_2^{b_1}$ to vary in the calculation of the
Bethe-Salpeter amplitudes and optimise a least-squares fit to the underlined
quantities in the table.  This simple expedient admits some of those
differences between the modulating functions observed in all of the more
sophisticated BSE studies and the possibility of remediating the separable
kernel's simplistic Chebyshev expansion.

An optimal fit is obtained with
\begin{equation}
\begin{array}{ccccc}
b_2^{+} & b_2^\pi & b_0^\rho & b_1^\rho & e_2^{b_1} \\
0.863 & 0.891 & 0.690 & 3.43 & -0.673,
\end{array}
\end{equation}
$(e_1^{b_1})^2+(e_2^{b_1})^2=1$, and yields the results presented in column
II, Table~\ref{tablea}, which has a rms relative error of 17\%.  (cf. 30\%
over five comparable items in Table~I of Ref.\cite{eric}.)  The most
significant effect of the relaxation is the change in sign and relative
magnitude of $e_2^{b_1}$.  Such a quantitative change is not too surprising
given that $\Gamma^{b_1}$ is the Bethe-Salpeter amplitude most affected by
shortcomings in the angular projection of the rank-2 separable Ansatz.  Thus
a satisfactory understanding of these phenomena is possible in our approach.
However, it is clear that a definitive study combining and extending
Refs.\cite{mr97,mt99} is required to conclusively answer the question posed
above.

We have applied a simple DSE model to the calculation of the spectrum and
decays of light axial-vector mesons.  Like light and heavy vector meson
observables\cite{mishaPRD}, they find a natural explanation in the
momentum-dependent dressing of quark propagators and the detailed form of the
meson Bethe-Salpeter amplitudes.  In the more sophisticated versions of this
approach these qualitatively important features are tied to the long-range
form of the interquark interaction, which in Refs.\cite{pankaj,mr97,mt99} is
represented via the infrared behaviour of ${\cal G}(k^2)$: the scalar
function that characterises the renormalisation-group-improved rainbow-ladder
truncation.  (The ultraviolet behaviour of $S$, $\Gamma_\nu$ and $\Gamma^H$
is model independent.)  

There are phenomena that cannot be described using the rainbow-ladder
truncation: $J^{PC}= 0^{++}$ meson masses and the $\eta$-$\eta^\prime$
complex being significant examples, and another the critical
exponents\cite{critexp} of the finite temperature chiral symmetry restoring
transition.  However, a qualitative understanding of the reasons underlying
these failures exists so that the truncation's domain of applicability is
becoming properly demarcated.  That allows for the sensible interpretation of
experimental results in terms of ${\cal G}(k^2)$, and thereby provides a tool
for the critical evaluation of contemporary numerical estimates\cite{Dmunu}
and, where required, improvements in the truncation of $K$, the
quark-antiquark scattering kernel.

We acknowledge helpful conversations and correspondence with M.A.~Pichowsky,
E.S.~Swanson and P.C.~Tandy.  Yu.L.K. gratefully acknowledges the hospitality
and support of the Physics Division at ANL during a visit in which some of
this work was conducted.  This work was supported in part by the Russian Fund
for Fundamental Research, under contract number 97-01-01040, and the US
Department of Energy, Nuclear Physics Division, under contract number
W-31-109-ENG-38, and benefited from the resources of the National Energy
Research Scientific Computing Center.  S.M.S. is a F.-Lynen Fellow of the
A.v. Humboldt foundation.


%
\begin{table}[t]
\caption{Calculated results compared with observed or inferred values, which
are taken from Refs.~\protect\cite{pdg,mishaPRD,isgur}.  We used the
calculated masses in Eq.~(\protect\ref{sepout}) with the exception of
$m_{h_1}=1.17\,$GeV and all dimensioned quantities are quoted in GeV.  Column
I: calculated using exactly those Bethe-Salpeter amplitudes obtained with the
separable Ansatz for ${\cal G}(k^2)$ in Eqs.~(\protect\ref{Ga}),
(\protect\ref{Gb}); Column II: optimised least-squares fit to the underlined
quantities obtained using a simple expedient that promotes further diversity
between the meson Bethe-Salpeter amplitudes.
\label{tablea}}
\begin{tabular}{lldd}
                                        & Obs.          & I     & II \\\hline
$\left.D/S\right|_{h_1\to \rho\pi}$  =: $R_{h_1}$  &                  & 
         0.81  & 0.25 \\
$\left.D/S\right|_{a_1\to \rho\pi}$  =: 
\underline{$R_{a_1}$}  & -0.1 $\pm$ 0.028 & 
        -0.092 & -0.075 \\
$\left.D/S\right|_{b_1\to \omega\pi}$  =: 
\underline{$R_{b_1}$} & ~0.29 $\pm$ 0.04  & 
         0.97  & 0.31 \\
$R_{a_1}/R_{b_1}$                       & -0.34 $\pm$ 0.11 &
        -0.095 & -0.25 \\
$R_{h_1}/R_{b_1}$                       &                 &
        0.84 & 0.83 \\
\underline{$g_{\rho\pi\pi}$} &               6.05 $\pm$ 0.02   & 
         9.58  & 8.18 \\
$\Gamma_{\rho\to \pi\pi}$              & 0.151 $\pm$ 0.001&  
         0.356 & 0.259 \\
\underline{$\Gamma_{a_1\to \rho\pi}$}      & 0.25 - 0.60      &     
         4.02 &  0.385 \\
\underline{$\Gamma_{b_1\to \omega\pi}$}    & 0.142 $\pm$ 0.009&  
         0.308 & 0.146 \\
$\Gamma_{h_1\to \rho\pi}$               & 0.360 $\pm$ 0.040&  
         0.573 & 0.301 \\
\underline{$f_{a_1}$}                   & 0.203 $\pm$ 0.018   &  
         0.121 & 0.221 \\
\underline{$f_{\rho}$}                  & 0.216 $\pm$ 0.005&  
         0.189 & 0.223 \\
\underline{$f_\pi$}                     & 0.1307 $\pm$ 0.0004 &  
         0.136 & 0.148 \\\hline
\end{tabular}
\end{table}
\end{document}